\documentstyle[preprint,pre,aps]{revtex}  
  
\begin{document}
\draft

\title{Random replicators with high-order interactions}
\author{ Viviane M.\ de Oliveira and J.\  F.\ Fontanari}
\address{Instituto de F\'{\i}sica de S\~ao Carlos,
 Universidade de S\~ao Paulo\\
 Caixa Postal 369, 
 13560-970 S\~ao Carlos SP, Brazil 
}

\maketitle

\begin{abstract}
We use tools of the equilibrium statistical mechanics of disordered 
systems to study analytically the statistical
properties of an ecosystem composed of $N$ species  
interacting via random, Gaussian interactions of order $p \geq 2$,
and deterministic self-interactions $u \geq 0$. We show that for 
nonzero $u$ the effect of increasing the order of the interactions
is to make the system more cooperative, in the sense that the 
fraction of extinct species is greatly reduced. Furthermore, 
we find that for $p > 2$ there is  a threshold  value which gives 
a lower bound to the concentration of the surviving species, 
preventing then  the existence of rare species  and,
consequently, increasing the robustness of the ecosystem to external
perturbations. 
\end{abstract}



Conservationists' arguments in favor of biodiversity have often 
appealed to the existence of intricate ties among apparently 
unrelated species in which,  for instance, the strengths of the 
interactions between any pair of species would depend on the
concentrations or frequencies of a variety of different ones 
\cite{Gaia}. Although 
the roles played by the total number of species, as well as by
the strengths of their pairwise interactions, in the stability of an
ecosystem  are now fairly well understood
both theoretically and experimentally \cite{Berlow,Ives,Morin},
it is still  not clear whether  high-order interactions among 
the species would 
actually bring any advantage, in the sense of a larger robustness, 
to the ecosystem. This is the main issue we address in this letter. 
Albeit the model proposed here is somewhat unrealistic from the 
biological viewpoint, since its dynamics is governed by a Lyapunov 
function, it clearly points out the advantages of high-order 
interactions, making clear-cut, nontrivial predictions such as the 
reduction of the number of extinct species and the existence of a 
concentration threshold which excludes rare species from  the 
ecosystem at equilibrium. 

Traditionally,  the study of co-evolution of species has been  
restricted to deterministic interactions \cite{hs-book}, however
the ever-present uncertainties about how the species are actually 
interacting allied to the overwhelming complexity of those 
interactions \cite{Morin} motivate an alternative, and perhaps 
complementary, approach in which  the strengths 
of the  interactions between the species are assigned at random. 
In this letter we solve analytically a model of co-evolution of $N$ 
species interacting  via random, high-order interactions. Our model 
is a generalization of the random replicant model 
\cite{Opper,Biscari,novo} which considers only pair interactions 
between the species.

We consider an infinite population (ecosystem) composed of 
individuals belonging to $N$ different species whose fitness 
${\mathcal F}_i ~(i=1, \ldots,N )$ are
the derivatives ${\mathcal F}_i = \partial {\mathcal F}/\partial x_i$
of the fitness functional ${\mathcal F}$ defined as
\begin{equation}\label{H_p}
- {\mathcal F} =  {\mathcal{H}}_p \left ( {\mathbf x} \right ) = 
u ~\sum_i x_i^p ~+
 \sum_{1 \leq i_1 <i_2 \ldots < i_p \leq N}
J_{i_1 i_2 \ldots i_p} \, x_{i_1} x_{i_2} \ldots x_{i_p}
\end{equation}
where $x_i/N$  is the concentration of species $i$. These variables
satisfy the constraints
\begin{equation}\label{constraint}
\sum_{i=1}^N x_i = N 
\end{equation}
and $x_i \geq 0 ~\forall i$. Here
the coupling strengths are statistically independent 
random variables with a Gaussian distribution
\begin{equation}\label{prob}
{\cal{P}} \left ( J_{i_1 i_2 \ldots i_p} \right ) =
\sqrt{\frac{N^{p-1}}{\pi p!}} \exp \left [ 
-\frac{ \left( J_{i_1 i_2 \ldots i_p} \right)^2 N^{p-1}}{p!} 
\right ] 
\end{equation}
for $i_1 < i_2 < \ldots < i_p$. The self-interaction parameter 
$u \geq 0 $ acts as a cooperation pressure limiting the growth of 
any single species, and it is crucial to guarantee the existence 
of a nontrivial thermodynamic limit, $N \rightarrow \infty$. 
It can be shown that the dynamics
\begin{equation}\label{rep_dyn}
\frac{d x_i}{dt} =  - x_i \left [ \frac{\partial {\mathcal H}_p }
{\partial x_i}
- \frac{1}{N} \sum_k  x_k  \frac{\partial {\mathcal H}_p }
{\partial x_k} \right ]
~ \forall i
\end{equation}
minimizes  $ {\mathcal H}_p \left ( {\mathbf x} \right )$ while
the mean $\sum_i x_i$ is a constant of motion (see, e.g., 
Ref. \cite{hs-book} page 240). This type of first-order differential
equation, termed  replicator equation,  has been used to describe
the evolution of self-reproducing entities (replicators) in a variety
of fields, such as game theory, prebiotic evolution and sociobiology,
to name only a few \cite{Schuster}. In particular,
a fourth-order interactions replicator equation was
shown to govern the  game dynamics in Mendelian (sexual) populations 
\cite{Mendel}.

For the sake of simplicity, in writing the fitness functional, 
Eq.\ (\ref{H_p}), we have implicitly assumed that the couplings 
$J_{i_1 i_2 \ldots i_p}$ are invariant under permutations of the 
indices $i_1, \ldots, i_p$. We must stress, however, that regardless
whether the couplings are  invariant or not, the interaction term in
the replicator equation, namely 
$\partial {\mathcal H}_p / \partial x_i$, will be invariant
to permutations of the species indices, and so the dynamics will
converge to a fixed point. In this sense, the  mere existence of a 
fitness functional (Lyapunov function) is a severe assumption  from 
the biological viewpoint. On the other hand, it allows full use of 
the tools of the equilibrium statistical mechanics to study 
analytically the properties of the fixed points of the corresponding 
replicator equation.

In the sequel we present the results of the replica analysis
of the statistical  properties of the ground state of
the multispecies interaction Hamitonian (\ref{H_p}).
Following the standard prescription of performing quenched averages
on extensive quantities only \cite{MPV}, we define the average
free-energy density $f$ as
\begin{equation}\label{f0}
- \beta f = \lim_{N \rightarrow \infty} \frac{1}{N} \left \langle 
\ln Z \right \rangle
\end{equation}
where 
\begin{equation}\label{Z0}
Z = \int_0^\infty \prod _j dx_j ~ \delta \left ( N - \sum_j x_j
\right ) \mbox{e}^{- \beta {\mathcal H}_p 
\left ( {\mathbf x} \right ) }
\end{equation}
is the partition function and $\beta = 1/T$ is 
the inverse temperature. 
Taking the limit $T \rightarrow 0$ in 
Eq.\ (\ref{Z0}) ensures that only the states that minimize 
${\mathcal H}_p \left ( {\mathbf x} \right )$
will contribute to $Z$.
Here $ \langle \ldots \rangle $ stands for the average over
the coupling strengths.
As usual, the evaluation of the quenched
average in Eq.\ (\ref{f0}) can be carried out through the
replica method \cite{MPV}. Within the replica-symmetric framework
we find that, in the thermodynamic limit, the average ground-state
energy per species is given by
\begin{equation}\label{E_0}
\epsilon_0 = \lim_{\beta \rightarrow \infty} f =
u \int_{-\infty}^\gamma Dz ~ x_s^p \left ( z \right ) - 
\frac{p}{2}~ y ~q^{p-1}
\end{equation}
where $Dz = dz \exp \left ( - z^2/2 \right )/\sqrt{2 \pi}$ is 
the Gaussian measure,
\begin{equation}\label{z_max}
\gamma = \left ( p - 2 \right ) q^{\frac{p-1}{2}} 
\left ( \frac{1}{u} \right )^{\frac{1}{ p-2 }}
\left ( \frac{y}{2} \right )^{\frac{p-1}{p-2 }}
\left ( \frac{p}{2} \right )^{\frac{ p} {2\left ( p-2 \right)}}
 - \Delta ,
\end{equation}
and $x_s \left ( z \right )$ is the positive solution of
\begin{equation}\label{x_s}
\frac{1}{2} p \left ( p - 1 \right ) y q^{p-2} x_s - p u x_s^{p-1} -
\left ( \frac{p}{2} q^{p-1} \right)^{1/2} 
\left ( \Delta + z \right ) = 0,
\end{equation}
which maximizes the effective Hamiltonian
\begin{equation}\label{Xi}
\Xi_x = \frac{1}{4} p \left ( p - 1 \right ) y q^{p-2} x^2 - 
u x^{p} - \left ( \frac{p}{2} q^{p-1} \right)^{1/2} 
\left ( \Delta + z \right ) x .
\end{equation}
We note that $x_s \left ( z \right ) = 0$ for $z > \gamma$. Here
the saddle-point parameters $q$, $y$, and $\Delta$ are given by 
the equations
\begin{equation}\label{sp1}
1 = \int_{-\infty}^\gamma Dz ~ x_s \left ( z \right ) ,
\end{equation}
\begin{equation}\label{sp2}
q = \int_{-\infty}^\gamma Dz ~ x_s^2 \left ( z \right ) ,
\end{equation}
and
\begin{equation}\label{sp3}
y ~ p \mid p - 1\mid = \int_{-\infty}^\gamma Dz ~ 
\frac{1}{\mid \frac{1}{2} y q^{p-2} - u  x_s^{p-2} 
\left ( z \right ) \mid } .
\end{equation}
Although in general these equations can be solved 
numerically only, we can easily obtain an analytical solution 
for large $u$:
\begin{eqnarray}
q & \approx & 1 + \frac{1}{u^2}~\frac{1}{2p \left ( p - 1 
\right )^2} \\
y & \approx & \frac{1}{u}~\frac{1}{p \left ( p - 1 \right )} 
\left [ 1 + \frac{1}{u^2}~\frac{1}{4 \left ( p - 1 \right )} 
\right ] \\
\Delta & \approx & -u ~ \left( 2p \right )^{1/2}  ~\left [ 1 - 
\frac{1}{u^2}~\frac{p+1}{4 p\left ( p - 1 \right )} \right ] .
\end{eqnarray}
The physical order parameter $q$ is defined by \cite{MPV}
\begin{equation}\label{q_mean}
q = \left \langle \frac{1}{N} \sum_i \langle x_i \rangle_T^2 
\right \rangle 
\end{equation}
where $\langle \ldots \rangle_T$ stands for a thermal average taken 
with the Gibbs  probability distribution
\begin{equation}\label{Gibbs}
{\mathcal W} \left ( {\mathbf x } \right ) = \frac{1}{Z} 
~\delta 
\left ( N - \sum_j x_j \right ) \; \exp \left [ - \beta
{\mathcal H}_p \left ( {\mathbf x} \right ) \right ] .
\end{equation}
Hence, values of $q$ of order of $1$ indicate the coexistence of a 
macroscopic number of species, while large values of $q$ signal the
dominance of a few species (i.e., the number of surviving species
increases like $N^x$ with $x < 1$) only. In Fig.~\ref{oliveira1} we present the 
physical order parameter $q$  as a function of the cooperation 
pressure $u$ for several values of $p$. As expected, for large $u$ 
the ecosystem is cooperative, in the sense that almost all species
survive, and so $q \approx 1$. For  small $u$ the system enters a
strongly competitive regime characterized by the divergence of
$q$, though the onset of this 
regime can  be postponed by increasing the order of the interactions
$p$, as illustrated in the figure. 
Interestingly, the analysis of
the effective Hamiltonian (\ref{Xi}) for $u=0$ shows that $x_s$ and
consequently $q$ [see Eq. (\ref{sp2})] are finite
only for $p < 1$,  which corresponds to a random version of
Szathm\'ary's model of parabolic growth \cite{Szat,wills}. 
As the divergence of $q$ signals the survival of
only a few species, the finitude of $q$ at $u=0$  is consistent with 
a parabolic growth for which the coexistence of all species is assured.  

To better understand the distribution of species in the ground state 
we calculate the distribution of probability that a certain species 
concentration, say $x_k$,  assumes the value $x$, defined by
\begin{equation}\label{prob_x}
{\mathcal{P} }_k \left ( x \right )
 =    \lim_{\beta \rightarrow \infty} \left \langle ~
 \int_0^\infty \prod_j dx_j  ~
\delta \left ( x_k - x \right )
{\mathcal W} \left ( {\mathbf x } \right )  \right \rangle
\end{equation} 
with ${\mathcal W} \left ( {\mathbf x } \right )$ given by 
Eq.\ (\ref{Gibbs}). As all species concentrations are equivalent we
can write ${\mathcal{P} }_k \left ( x \right ) = 
{\mathcal{P} }\left ( x \right ) \forall k$. Moreover,
to handle a possible singularity in the limit 
$\beta \rightarrow \infty$ it is more convenient to consider instead
the cumulative distribution function 
%
${\mathcal{C} }\left ( x \right )  = 
\int_0^{x} dx' \, {\mathcal{P} }\left ( x' \right )$ .
%
Carrying out the calculations we obtain
\begin{equation}\label{cum}
{\mathcal{C} }\left ( x \right )  = 
\int_\gamma^{\infty} Dz + \int_{-\infty}^\gamma Dz ~
\Theta \left [ x - x_s \left ( z \right ) \right ] 
\end{equation}
where $\Theta (x) = 1$ for $ x > 0$ and $0$ otherwise.
For $u \rightarrow \infty$ we find 
${\mathcal{C}}(x) = \Theta \left ( x - 1 \right )$ regardless of
the value of $p$ since in this case the equilibrium solution is 
$x_i = 1~\forall i$. An interesting feature of the cumulative 
distribution function is that ${\mathcal C } \left ( 0 \right )$ 
is nonzero, indicating thus that the probability distribution 
${\mathcal{P} }\left ( x \right )$ has a delta peak at $x = 0$. 
In fact, the first term of the rhs of Eq.\ (\ref{cum}), i.e.,  $
{\mathcal C } \left ( 0 \right )$,
yields the fraction of extinct species in the ground state.
Moreover, as shown in  Fig.~\ref{oliveira2}, the constancy of 
${\mathcal{C}}(x)$ up to a threshold concentration value 
$x_t = x_s \left ( \gamma \right )$ indicates that there is a lower 
bound to the concentration of any surviving species. As illustrated 
in Fig.~\ref{oliveira3}, $x_t$ decreases with the cooperation pressure
$u$, and increases with the order $p$ of the interactions. Since 
$x_t =0$ at any finite value of $u$ for $p=2$, a nonzero value of 
$x_t$ is an effect of  the higher-order of the interactions. Clearly, 
the nature of this concentration threshold is totally distinct from 
that of the threshold obtained in the limit $u \rightarrow \infty$,
which equals $1$ for all $p$. As expected, $x_t \rightarrow \infty $ 
for $u=0$ since ${\mathcal C } \left ( x \right ) = 1$ for all $x$ 
in this limit, while 
\begin{equation}
x_t \approx \left ( 2 p u^2 \right )^{-1/\left ( p-2 \right)} 
\end{equation}
for $u$ large. Of course, the existence of such a
threshold has far-reaching consequences on the stability and
robustness of the population against external perturbations since it
implies the absence of rare, and hence proner to extinction, species 
whose loss might cause dramatic effects in the whole population 
\cite{Berlow}. Furthermore, for fixed $u$ the fraction
of extinct species decreases with increasing $p$ 
(see Fig.~\ref{oliveira4}). More pointedly, for large $u$ we find
\begin{equation}
{\mathcal C } \left ( 0 \right ) \approx 
\frac{1}{\left ( 4 \pi p \right )^{1/2} }~\frac{1}{u}~
\exp \left ( - p u^2 \right ) , 
\end{equation}
which shows that increasing the order of the interactions among the
species makes the ecosystem more cooperative, thus corroborating  the 
conclusions drawn from the analysis of  Fig.~\ref{oliveira1}. 

We have verified the validity of the replica-symmetric solution by 
performing the standard stability analysis \cite{AT}.  In particular, 
that solution becomes unstable 
for $u$ smaller than $1/\sqrt{2} \approx 0.707$ and $ 0.106 $ for 
$p=2$ and $3$, respectively, while for $p \geq 4$ we find that it 
is unstable only at $u=0$. Hence our main results are not
affected by the (local) instability of the replica-symmetric solution.

To conclude we must emphasize that our results describe the 
equilibrium properties of the population only. Important issues such
as whether the absence of rare species at equilibrium would imply
that the ecosystem is stable with respect to the invasion of rare 
mutant species, or whether the effect of a perturbation decreasing  
the concentration of a single species to a value below $x_t$ would 
lead to the collapse of the entire ecosystem,
can be addressed only through a dynamical approach  \cite{novo}, 
which is beyond the scope of our present work. We hope the nontrivial 
predictions of our model will provide motivation for the proposal of 
more realistic models of high-order multispecies interactions.

\bigskip

\bigskip

We thank Peter F. Stadler and Rita M. Z. dos Santos for useful 
discussions. The work of J.F.F. is supported in part by Conselho
Nacional de Desenvolvimento Cient\'{\i}fico e Tecnol\'ogico (CNPq)
and  Funda\c{c}\~ao de Amparo \`a Pesquisa do Estado de S\~ao Paulo 
(FAPESP), Proj.\ No.\ 99/09644-9. V.M.O. is supported by FAPESP.



\begin{figure}
\caption{Physical order parameter $q$ 
as a function of the cooperation pressure $u$
for $p= 2$, $3$, $5$ and $10$.}
\label{oliveira1}
\end{figure}

\begin{figure}
\caption{Cumulative distribution function of the ground-state
species concentrations for $p=3$ and several values of $u$ as indicated
in the figure. The dashed line is the result for $u \rightarrow 
\infty$.}
\label{oliveira2}
\end{figure}

\begin{figure}
\caption{ Concentration threshold $x_t$ as a function of $u$ for $p=3$,
$5$, $7$, $9$, $11$ and $13$. Note that $x_t = 0$ for $p=2$.}
\label{oliveira3}
\end{figure}

\begin{figure}
\caption{Fraction of extinct species in the ground-state $C(0)$ as a 
function of $u$ for several values of $p$ as indicated in the figure.
Note that $C(0)=1$ for $u=0$ indicating that only a few 
species survive in the thermodynamic limit.}
\label{oliveira4}
\end{figure}


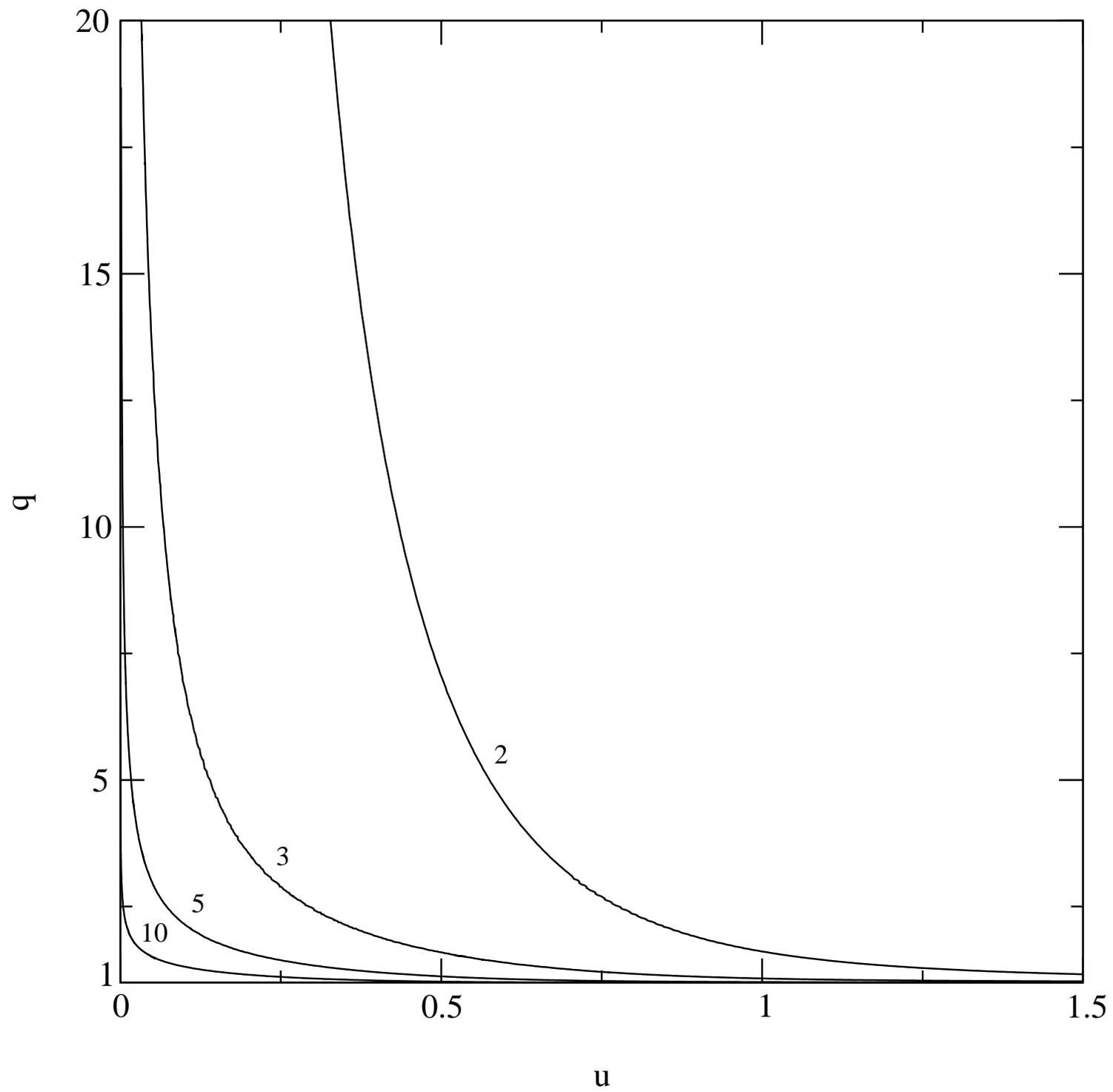
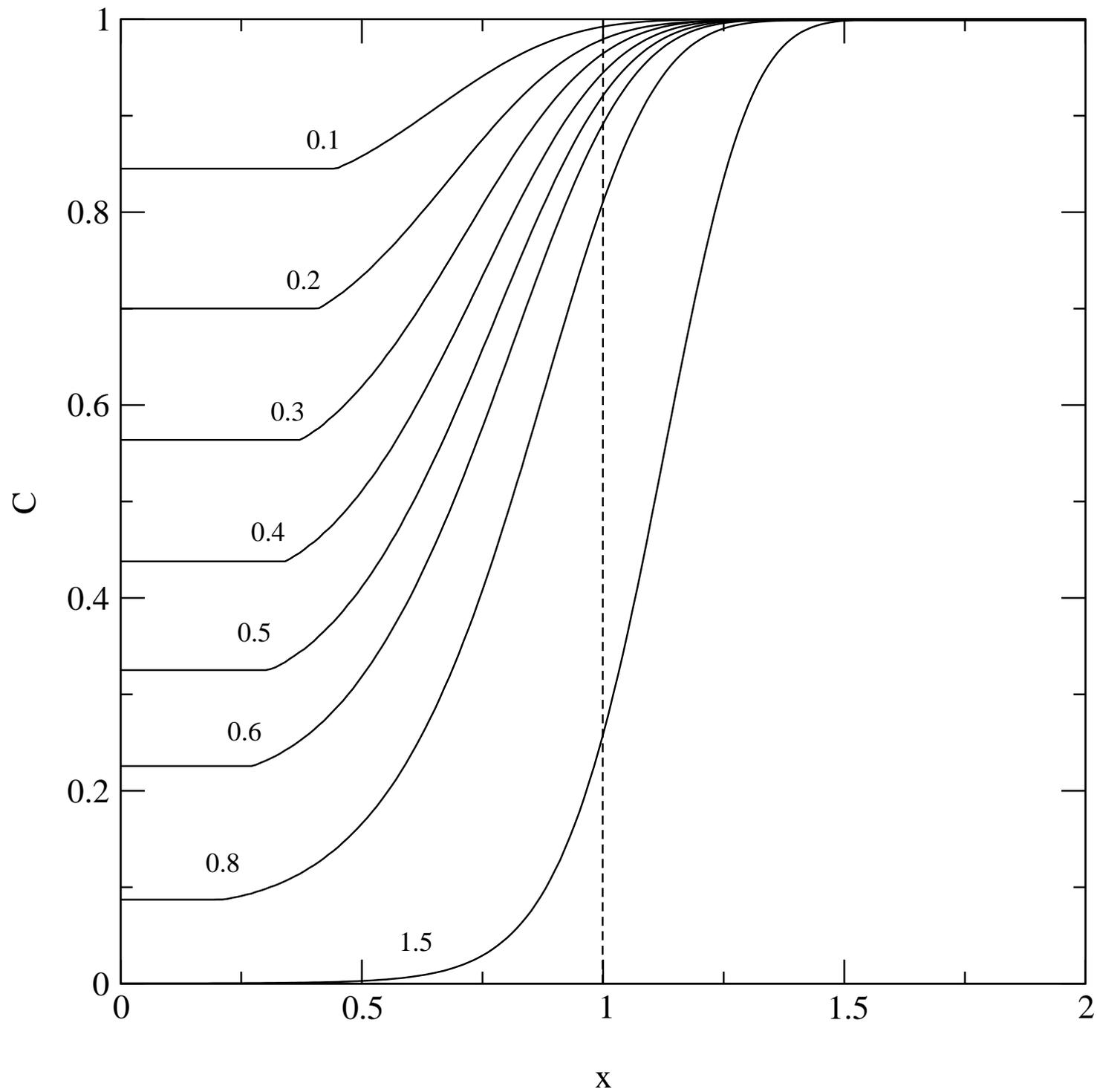
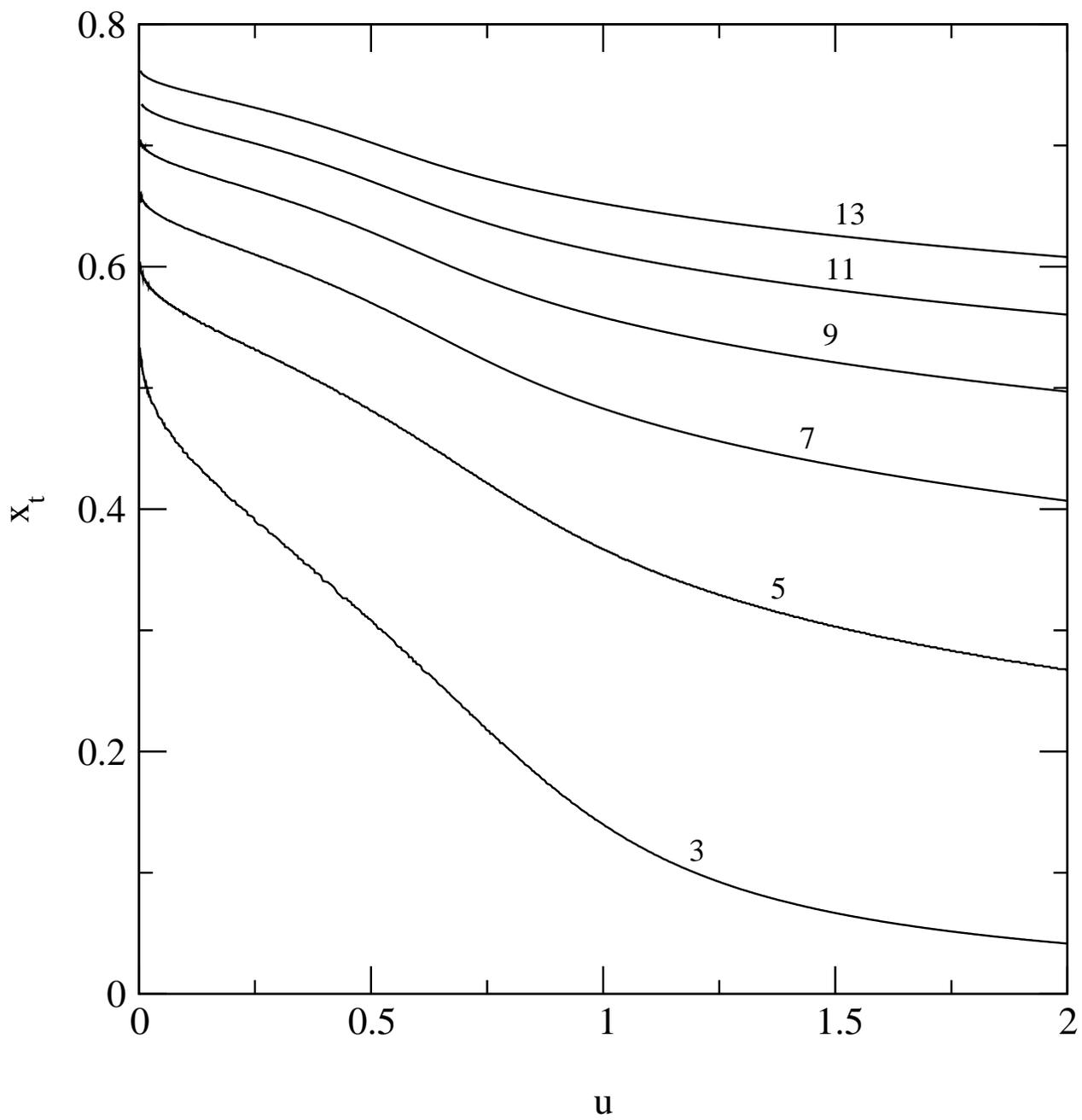
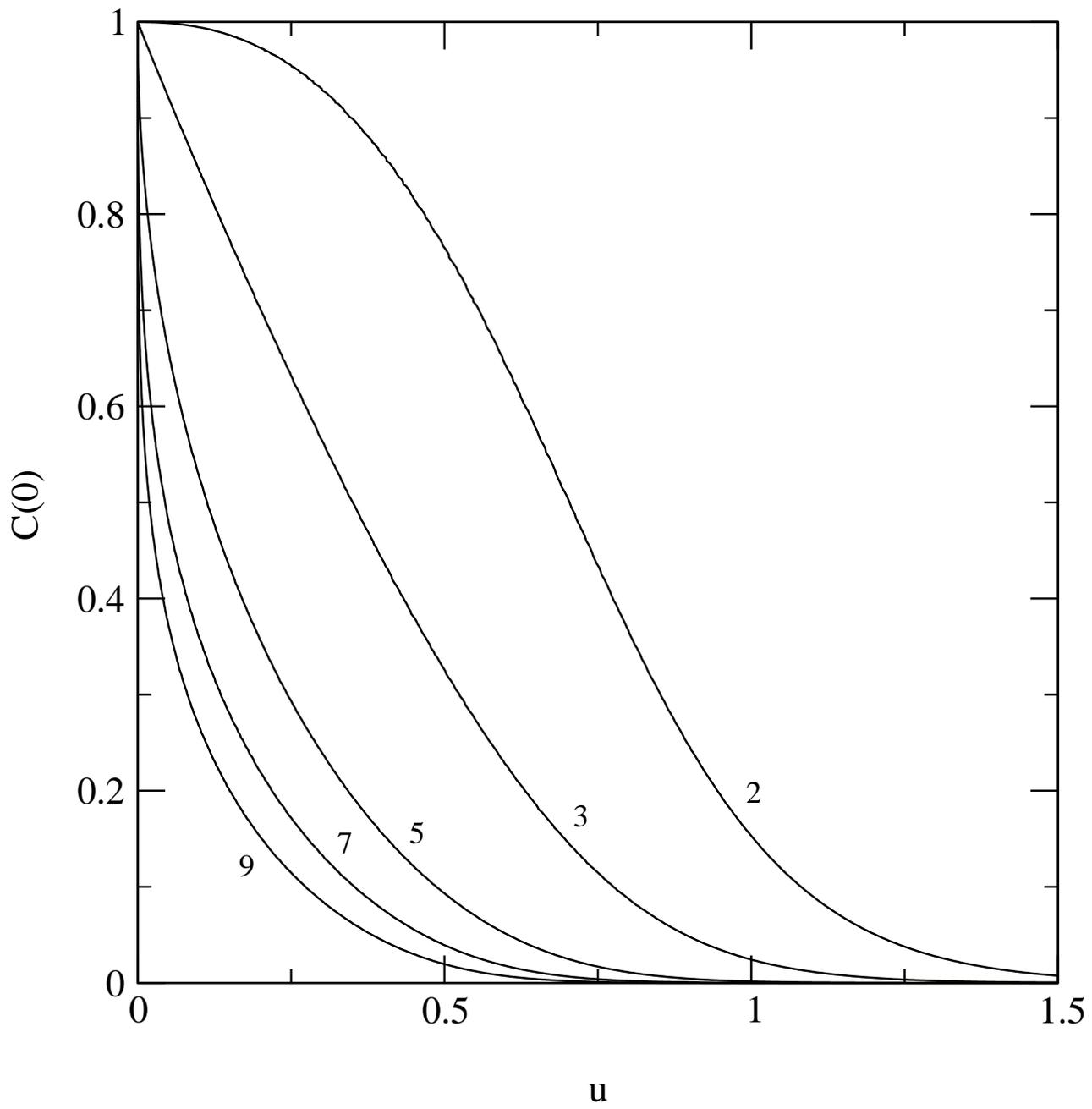
\end{document}